\documentclass[12pt]{article}
\usepackage{amsmath}
\usepackage{amsfonts}
\usepackage{amssymb}
\usepackage{graphicx}

\ifx\pdfoutput\@undefined\usepackage[usenames,dvips]{xcolor}
\else\usepackage[usenames,dvipsnames]{xcolor}
\IfFileExists{pdfcolmk.sty}{\usepackage{pdfcolmk}}{} 
\fi
\usepackage[plainpages=false,pdfpagelabels,pagebackref=false,naturalnames=true,hyperindex=true,pdftitle={A New Kind of Finance},pdfauthor={Philip Z. Maymi}]{hyperref}
\hypersetup{colorlinks=true,
urlcolor=Cerulean,linkcolor=BrickRed,citecolor=RoyalBlue,a4paper,
  pdfpagemode=None,
  pdfstartview=FitH}
\usepackage[all]{hypcap}

\begin{document}

\title{A New Kind of Finance}
\author{Philip Z. Maymin\\NYU-Polytechnic Institute, USA.
}
\date{}

\maketitle

\begin{abstract}
Finance has benefited from the Wolfram's \textit{NKS} approach but it can and will benefit even more in the future, and the gains from the influence may actually be concentrated among practitioners who unintentionally employ those principles as a group.\\

\end{abstract}

The insights and techniques from Stephen Wolfram's \textit{A New Kind of Science} \cite{Wolfram}---namely that simple systems can generate complexity, that all complexity is maximal complexity, and that the only general way of determining the full effects of even simple systems is to simulate them--are perhaps most useful, and least applied, in the field of finance. 

The influence of \textit{NKS} on the current state of finance depends on the particular area of finance being studied. In the area of market-based finance, a unique minimal model of financial complexity has been discovered. In the area of government-based finance, the same minimal model has been used to test the effects of different regulatory regimes. Those are academic results; finance is ultimately a practitioner's field. From the perspective of practitioners, a result linking computational efficiency and market efficiency has been found.

In short, finance has benefited from the \textit{NKS} approach but it can and will benefit even more in the future, and the gains from the influence may actually be concentrated among practitioners who unintentionally employ those principles as a group.

What is finance, anyway? It can be hard enough to pronounce, let alone define. Should it be \textit{FIE-nance} , or \textit{fih-NANCE}? I've studied, researched, and practiced in the field for most of my life, and I still don't know how to pronounce it. Fortunately, I'm not alone. Dictionaries list both as acceptable pronunciations.

Perhaps it depends on whether the word is used as a verb or a noun. After some prodding, English American speakers will usually agree that the former is the proper pronunciation for the verb form, as in when you \textit{FIE-nance} a car, and the latter for the noun form, as in when you protest the bailouts of companies involved in high\textit{fih-NANCE}. The British feel just as strongly that one form is a verb and the other a noun---but the opposite ones.

%\noindent\(\textbf{Speak}[\textbf{``finance vs. fenance, from fine and fin''}];\)\\

The term originated from the French \textit{fin}, marking the end of a contract through the fulfillment of an obligation or debt. As such, finance has a noble libertarian heritage. But it shares the same root as the unfortunately authoritative English word ``fine'', meaning a penalty payment to a government.

This pronunciation ambiguity is not just a curiosity. It is an omen and a symptom of the deep divide in the study of finance between market-based approaches and government-based approaches. And it turns out that this deep divide explains why in finance the \textit{NKS} perspective is both so sorely needed and so often neglected.

\section{Market-Based Approaches}

Markets resulting from voluntary trade tend to be complex phenomena. A typical price chart shows wild swings, big jumps, bubbles, and crashes. These are even more obvious when we look at the chart of returns instead of prices. (Recall that the return from one day to the next is the percentage you would have earned if you bought it one day and sold it on the next.)

\begin{figure}[htb!]
  \centering
\includegraphics[scale=1.2]{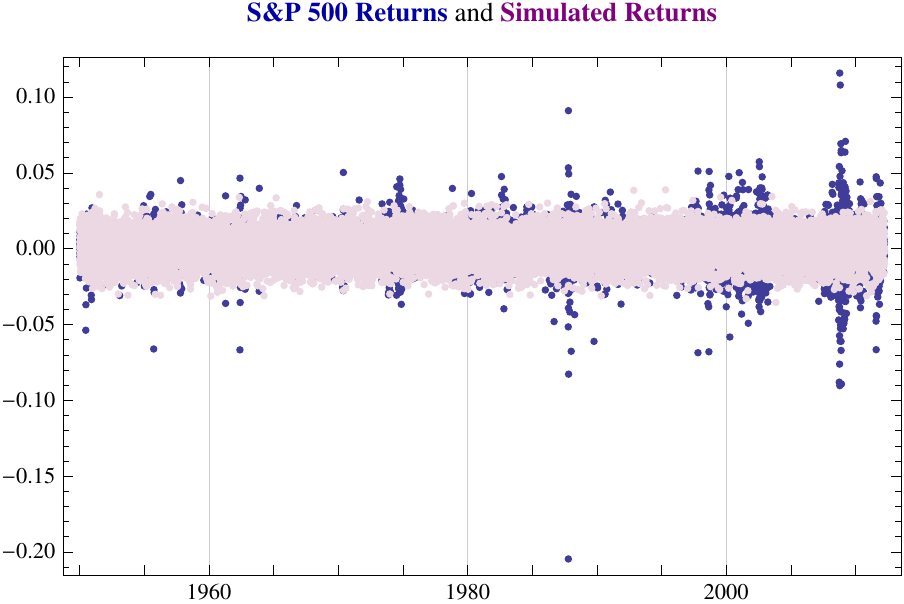}
\caption{The dark dots are the actual daily returns of the Standard $\&$ Poor's 500, the most widely followed broad based U.S. market index. The light dots overlaid on top are simulated returns from the Normal distribution having the same mean and standard deviation as the actual returns. You can see that the blue dots vary wildly, much more than could be expected from a Gaussian distribution. In addition, these periods of higher volatility\index{volatility} tend to cluster together. And finally, there is Black Monday, October 19, 1987, when the market fell by 20 percent.}
\end{figure}

As Wolfram has noted, most academic market-based approaches to explaining or understanding these complexities essentially ignore the vast amount of seeming randomness and focus on the few pockets of predictability. For example, momentum, the idea that winners will keep winning and losers will keep losing, seems to be a persistent feature of many markets, and has been the subject of thousands of scholarly papers after its first documentation by Jegadeesh and Titman \cite{Jegadeesh}. But the effect of momentum, while profitable, is still rather small compared to the vast degree of randomness.

Virtually the only tools used for this standard strand of research are regression analysis, attempting to explain individual security or portfolio returns through a fixed number of factors, and portfolio construction, attempting to sort portfolios into buckets based on some factors or indicators and explore the difference in future performance between the highest and the lowest buckets. 

In \textit{NKS}, Wolfram explored the alternative approach of trying to model the randomness directly rather than ignoring it. He proposed a one-dimensional cellular automaton model where each cell represents an agent's decision to buy or sell, and the running totals of black cells can be used to infer a market price. Jason Cawley\index{Cawley, J.} has generalized this model in a \textit{Mathematica} demonstration. 

In a sense, cellular automata\index{cellular automata} models for financial prices are a subset of the more general recent approach of agent-based modelling.\index{agent-based modelling} Here, agent behavior is programmed into several varieties, initial proportions of each are chosen, and the interactions between those agents generates market transactions and prices. Gilbert \cite{Gilbert} offers a comprehensive introduction and treatment of this literature. The Santa Fe Institute created an artificial stock market a few decades ago; Ehrentreich \cite{Ehrentreich} focuses on agent-based finance and specifically on the lessons of this market. The ability to create multi-agent models has become even easier with the introduction of specialized environments for such tasks such as NetLogo \cite{Wilensky}.\index{NetLogo}

However, all such agent-based models, including Wolfram's, rely on multiple agents interacting and trading with each other, often with multiple securities too. In the spirit of \textit{NKS}, we should ask: could a single representative investor trading\index{trading} a single security generate complexity?

This was exactly the question I asked during the NKS Summer School of 2007. I realized that because there was no one for the lonely representative investor to trade with, and no other assets for him to compare his to, he would have to be a technical trader, someone who makes decisions based solely on the past history of prices. Technical traders are also called chartists because they often rely on graphical representations of past prices, such as when moving averages of different lookback windows cross, or when the prices seem to form a recognizable visual pattern. Indeed, given the recognition in \textit{NKS} that our natural visual ability was well-adapted to discerning complexity, it seemed reasonable to assume that some of the skills of a technical trader could possibly result in complexity in the price series directly.

Although technical traders can rely on any function of historical prices, a simpler and yet still fully general approach would be to model a trader as evaluating an arbitrary algorithm taking as input the previous prices, or price changes, or even just the signs of those price changes, starting with the most recent first. 

The primary benefit of the NKS Summer School is working one-on-one with the author. Indeed, Wolfram suggested using an iterated finite automaton (c.f. Wolfram \cite{Wolframessay}) to model the internal algorithm of the trader. An iterated finite automaton (IFA) \index{iterated finite automaton} takes one list of symbols and outputs another, and can have internal states. It is thus a collection of rules of the form: 

\begin{center}
\noindent \textbf{\{state1,input\}} $\to$ \textbf{\{state2,output\}}
\end{center}

Trivially, no single-state IFA generates complexity. Among all of the 256 possible two-state IFAs, there turned out to essentially be only one unique trading rule that generated some form of complexity. Using Wolfram's IFA numbering scheme, this was trading rule 54, depicted by the graphical network below.

\begin{figure}[htb!]
  \centering
  \includegraphics[scale=.9]{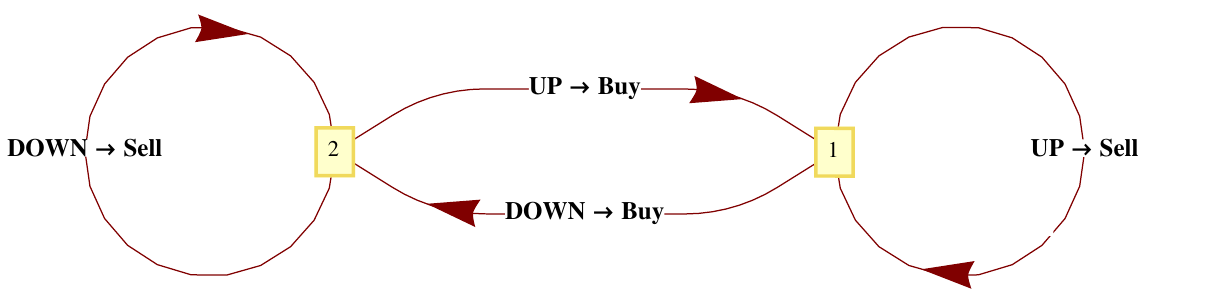}
\caption{The boxes represent the two internal states of the trader's rule. He starts every day in state 1. He looks back at the $n$ most recent price changes, starting with the most recent, and follows the arrows until he reaches the $n$th-most recent one. }
\end{figure}

Suppose his lookback window is $n = 3$ days. If today is Thursday, then he would have to look back at the market price change on Wednesday, Tuesday, and Monday, in that order. Let's say the market was down Wednesday. So he leaves state 1 following the DOWN arrow, which leads him to state 2. That DOWN arrow also outputs a Buy signal. This can be viewed as his current thinking on what to do in the market, but it is not his final decision because he has not looked at all of the past few days that he intended to. 

Next he would need to look at Tuesday's price change. Suppose it too was down. Then he would follow the DOWN arrow out of his current state, state 2. This arrow leads him back to state 2, and updates his current thinking to Sell. 

Ultimately, his decision on whether to buy or sell will now depend on what the price change on Monday was: if the market had been down, he would now sell, and if it had been up, he would now buy. Whatever he does, the market follows, because he is the representative investor. So if he were to Buy, no transactions would actually take place, because he has no one to trade with, but the level of the market would go up so he is now indifferent about buying more. The next day, he starts the process all over again, starting with the most recent price change, which happened to be up.

This rule 54 generates quite complex behavior, for virtually any lookback window. The graphs below show the price processes for a variety of lookback windows (See Fig.~3).

\begin{figure}[htb!]
  \centering
\includegraphics[scale=1]{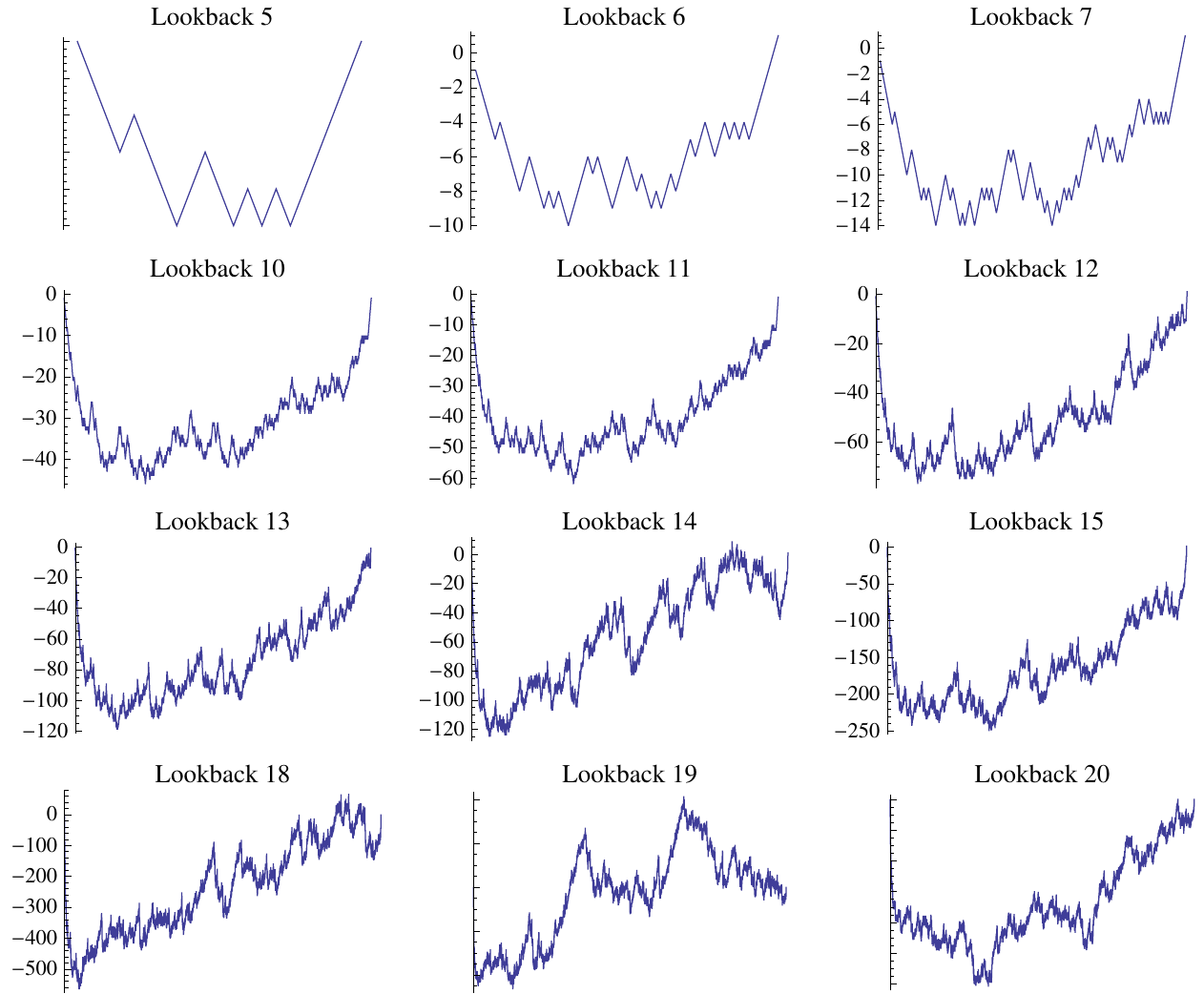}
\caption{Notice how the prices jump down drastically before coming back up. Because the prices are always deterministically calculated, they will eventually cycle, and could in principle start anywhere along the cycle. Thus, the big jump down could occur later in the cycle. Furthermore, the lookback window can be made larger so that the cycle time is longer than the age of the universe: looking back just 22 ticks means the cycle time is more than four million ticks.}
\end{figure}

So it is true that the absolute simplest model of trading can indeed generate complex price patterns, validating the key insights of \textit{NKS} and finally answering a question that Wolfram had worked on for decades. With just a single trader and a single asset, and only two internal states, there is essentially a unique rule that generates complex security prices. This is the minimal model of financial complexity (Maymin, \cite{Maymin2011a}).

But how complex is the generated price series? We have seen above that real markets suffer from many irregularities. Specifically, the stylized facts about market returns relative to independently distributed Normal returns are that real returns have higher kurtosis (fatter tails), negative skewness (more extreme jumps down), and a rich panoply of autocorrelations (generating mean reversion or momentum at different horizons).

By taking a lookback window of $n=22$ and partitioning the up and down ticks into buckets large enough to interpret their rolling sum as a daily return, we can estimate the implied kurtosis, skewness, and correlations of the resulting price series. Surprisingly enough, it turns out that all of the troublesome stylized facts of real markets occur in the generated price series as well!

Thus, the unique, simple, and minimal model of financial complexity, with no parameters to tweak, serendipitously ends up explaining much of what
we see in real markets.

What does the rule do, exactly? Do such traders exist? In general, there need be no easy description of a trading rule. But in this case, there happens to be a very simple explanation. Notice that state 1 is an UP-absorbing state: any UP day will bring the trader to state 1. Similarly, state 2 is a DOWN-absorbing state. Thus, rule 54 ultimately merely compares the two earliest days of its lookback window: if the price change $n - 1$ ticks ago were the same as the price $n$ ticks ago, then the investor would sell; if they were different, he would buy.

An alternative interpretation is that the representative investor look each tick and compares it to the previous one. If they are the same, whether both up or both down, he sells; if they are different, either up and then down or down and then up, he buys. However, his order does not take effect immediately but rather experiences a delay of $n-1$ ticks. Put this way, such a trading rule can be expressed as a combination of four
commonplace rules: profit taking in bull markets, momentum in bear markets, buying on dips, and buying on recoveries. 

Naturally, the minimal model can be extended to multiple states, multiple assets, and multiple traders, and complexity again emerges, with more variety as well. But it is interesting that even the minimal model is able to fit actual returns so well, and so much better than random walks or Brownian motions, the standard assumptions of non-NKS-influenced finance.

Clearly, the \textit{NKS} approach is useful in market-based finance. So why is it not more frequently used in academic circles? The reason is selection bias.

The bane of academic financial research is selection bias. Selection bias in data can falsely suggest that certain assets or industries had high expected returns, only because those were the only ones who survived long enough to be in the dataset. Selection bias may even be latent and quite subtle: one of the longest puzzles in finance is the equity premium puzzle documented by Mehra and Prescott \cite{Mehra} in 1985, noting that historical average returns have been far too high to be explained by risk aversion, the standard explanatory tool of financial economics. But we will never know if the selection bias of having had a booming stock market for many decades is what allowed us the luxury of asking why have our stock returns been so large.

But by far the biggest concern is selection bias of the models, also known as data snooping. If we posit a model that is influenced by what we have seen, then tests of the model are contaminated. At the extreme, you can always optimize the parameters of any family of models to get the best possible fit, but you will never know if you are not just overfitting noise.

Partially in an attempt to combat this problem, and partially because finance is often viewed as a discipline of economics, academic literature in the area is virtually required to motivate any analysis with detailed reasoning why the model makes sense \textit{a priori}. Of course, it is impossible to tell by reading a paper whether the model indeed was formulated prior to any observation of the data or whether it was retrofit onto it later, or, less obviously, whether it was just the lucky one of many models tested that happened to work. Academics rarely (though not never) publish the results of failed models.

This attachment to motivation is the biggest hurdle to wider acceptance of the useful tools and techniques of the \textit{NKS} framework. Mining the computational financial universe requires abandoning all preconceptions of what should or should not work and instead trying hundreds, thousands, millions of possibilities to see what does indeed work. By the Principle of Computational Irreducibility,\index{Principle of Computational Irreducibility} the motivation game can not work in general, and can even be a hindrance to the truth. The \textit{NKS} approach to market-based finance requires overcoming enormous inertia to flip standard academic practice completely on its head. 

That's a tough row to hoe, but there have been some other inroads. Explicitly, Zenil and Delahaye \cite{Zenil} investigate the market as a rule-based system by comparing the distributions of binary sequences from actual data with those resulting from purely algorithmic means. On a more implicit level, many otherwise standard-seeming financial results seem to be more willing to test literally all possible strategies or combinations, reserving their motivation and justification only to the form of the model. The tide may not have started to turn yet, but the waves are starting to froth.

\section{Government-Based Approaches}

Markets resulting from government fiat tend to be simple price fixings. Even the ostensibly more general price floors or ceilings end up being price fixings anyway because otherwise the legislation is useless. So a time series of government-controlled prices tend to look like a constant, experiencing nearly zero volatility... until the government can no longer control the price and the pent-up volatility explodes all at once. Imagine a currency peg about to break or the stock market hitting an automatic circuit breaker curbing trading. When trading resumes, the true price will likely be very different from the most recently reported price. 

In exploring regulatory issues and their possible effects on markets, there are two traditional approaches: theoretical and econometric. The theoretical approach solves for the equilibrium in a particular standard model and evaluates how it changes under different regulatory regimes. The econometric approach attempts to analyze past regulatory changes to isolate the effects of unanticipated regulatory changes. These two approaches sometimes agree and sometimes disagree, and each has its own pitfalls.

A unifying way of viewing both approaches is to observe that they each effectively assume a particular process for the evolution of market prices, and then translate regulatory changes into different values for the particular parameters. Theoretical models attempt to solve for what the new parameters will be while the econometric models attempt to estimate them from the historical record.

There is a third way, the \textit{NKS} way: one could use a rules-based approach with regulatory overrides. Specifically, one could imagine the rule 54 trader wanting to sell the asset but being stopped by government forces intent on propping up the market.

This is now a question of computational search. It is unclear ahead of time what the effect will be. The best way to find out is to simulate it. In Maymin \cite{Maymin2009}, I did just that, and showed that regulation in general makes market price processes appear to be more Normal and less complex (until, of course, the regulation can no longer be afforded). Particular periods, however, could actually appear even worse than the non-regulated version.
Further, the results from pricking bubbles and propping up crashes are not symmetrical: specifically, if regulations were to prick apparent bubbles, then propping up apparent crashes makes no additional difference (see Fig.~4).

\begin{figure}[htb!]
  \centering
\includegraphics[scale=.76]{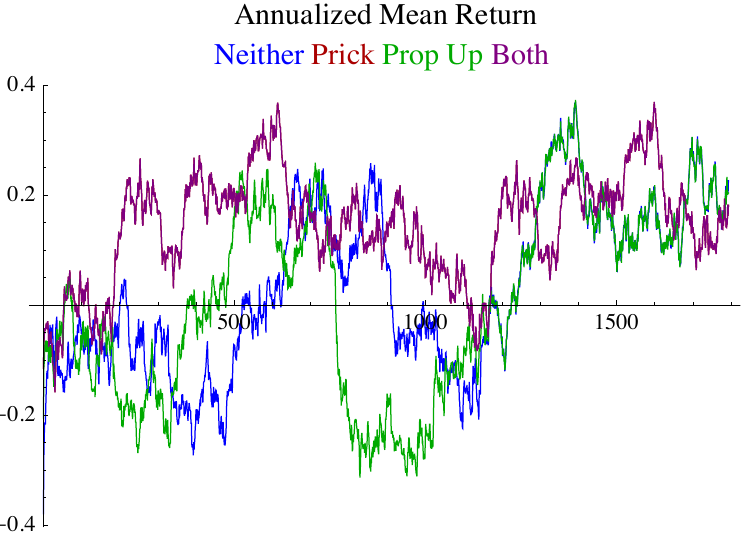}\includegraphics[scale=.76]{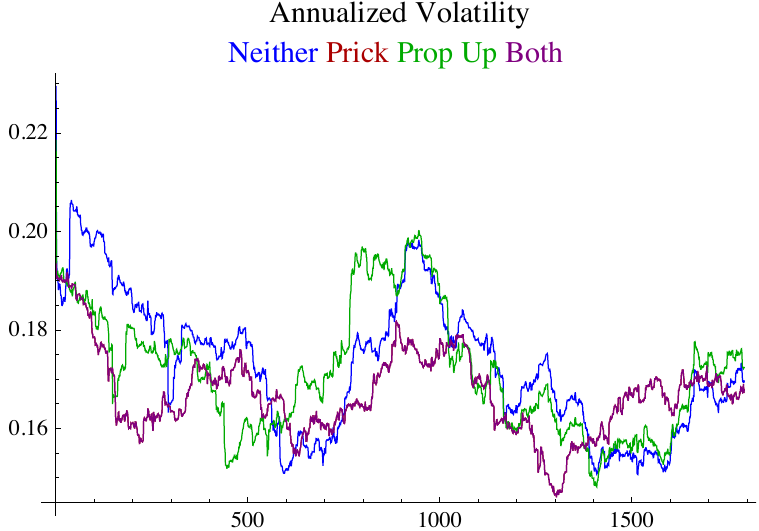}\\\includegraphics[scale=.76]{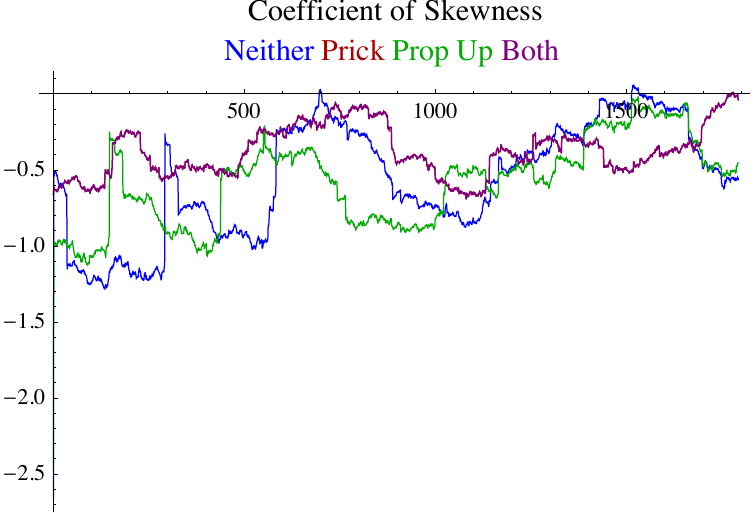}\includegraphics[scale=.76]{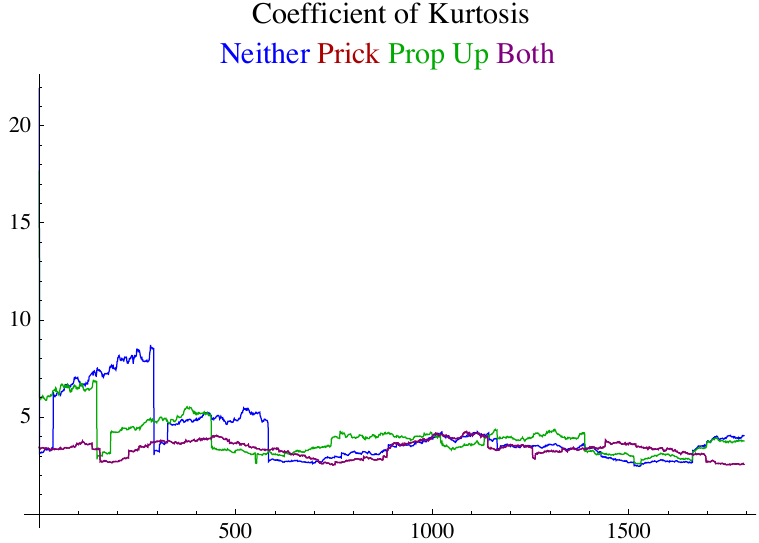}
\caption{The graphs  show rolling moment estimates from these four different regulatory regimes.}
\end{figure}

An even more direct result can be found in Maymin and Lim \cite{Maymin2012} where we compare regulations directly on a cellular automaton model. In the context of environmental regulations, suppose each cell represents an entity that can choose whether or not to pollute. And suppose the rule governing whether you pollute or not depends entirely on what you and your neighbors did in the previous instance. For concreteness, let's say it is Wolfram rule 110, which he has shown to be computational universal, or maximally complex (see Fig.~5). 

\begin{figure}[htb!]
  \centering
\includegraphics[scale=.9]{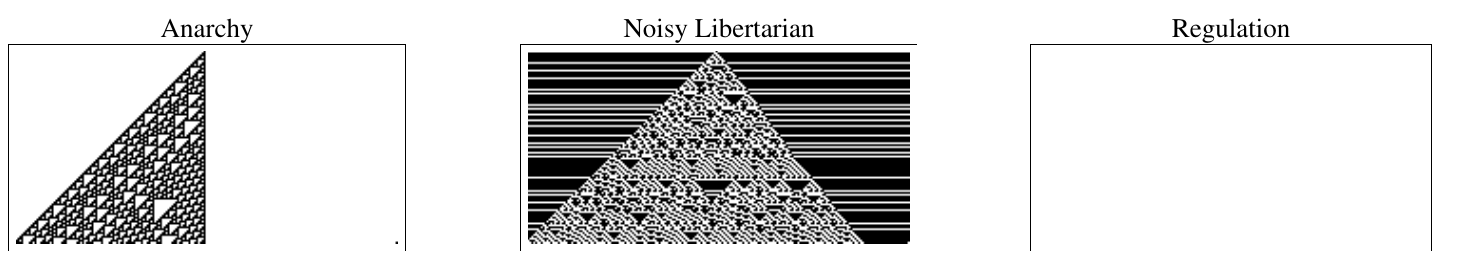}
\caption{Call anarchy the state of no overriding law, neither \textit{a priori} regulation nor \textit{ex post} justice. Then the half of the people on the right hand side never pollute, while those who occasionally pollute exhibit interesting, indeed maximal, complexity. Under complete \textit{a priori} regulation, no one would pollute ever, leading to zero complexity. But under \textit{ex post} albeit noisy justice in which with some probability those who polluted last time will now be polluted on by those who had abstained, maximal complexity is restored. Furthermore, even that half of the population that would not have polluted under anarchy now does occasionally pollute. Bearing in mind that pollution is a cost with associated benefits, and that some amount of pollution is likely to be optimal, we can draw conclusions about which system accomplishes what we want.}
\end{figure}

With the \textit{NKS} approach to regulation in general, both financial and otherwise, we are able to see the effects of varying kinds of regulatory overrides on top of a simple system of otherwise static rules. I expect that for the government-based strand of finance research, the \textit{NKS} approach will eventually come to dominate the field, as it represents the only way I can see of performing true experiments on the possible effects of different proposed regulations.

\section{Practitioners}

While academics and regulators play a loud part in finance, the silent super-majority are practitioners: traders, investors, and speculators who have a vested interested in keeping quiet and keeping secrets. Practitioners do not care how to pronounce the word ``finance,'' and they switch randomly from one to the other. They represent by far the most important constituency. Can an \textit{NKS} approach help them too?

In one sense, they represent the heart of the \textit{NKS} approach. Markets are complex but by the Principle of Computational Equivalence they are no more complex than other maximally complex things. Complex things can often be modeled by simple rules. When even the simplest of rules constitute an astronomical number of possibilities, the only possible approach, by the Principle of Computational Irreducibility, is exhaustive or random search. Thus, together, practitioners are essentially mapping and mining the financial computational universe, even if they are doing so unintentionally and occasionally redundantly.

It turns out that this task of finding a profitable strategy in past prices is one of the hardest computational problems on the planet. Indeed, I have shown that this task is as hard as solving satisfiability or the traveling salesman problem. In other words, markets will be efficient---that
is, there will be no profitable trading strategies based on past prices because they would have all been discovered and exploited---only if all other difficult problems have also been solved.

Surprisingly enough, I have also shown the converse: that if the markets happen to be efficient, then we can actually use those markets to solve
the other difficult problems. We can, in effect, ``program'' the market to solve general computational problems.

Thus, market efficiency and computational efficiency turn out to be the same thing. This paper, Maymin \cite{Maymin2011b}, sparked the creation of \textit{Algorithmic Finance}, a new journal and indeed a new field launched specifically to continue the insights from merging computational efficiency and
market efficiency. I am the managing editor of the journal and Stephen Wolfram is on the advisory board. With this journal, we hope to continue the journey of exploring \textit{NKS}-inspired approaches to the field of finance. 

\section{Conclusions}

The insights from \textit{NKS} are general, deep, and broad: simple rules can generate complexity; beyond a small threshold, all complexity is maximal
complexity; the only way of evaluating even simple systems that generate complexity is to run them and see. In finance, these insights are critical
for understanding markets and their evolution, particularly as trading moves ever closer to complete automation.

Both the journal \textit{Algorithmic Finance} and the field of algorithmic finance rely on these insights to grow. Applications as varied as high frequency finance and automated trading, the heuristics of behavioral investors, news analytics, statistical arbitrage, and dynamic portfolio management
all reside at the intersection of computer science and finance, and could, and have, and will continue to benefit from the tools of \textit{NKS}.

\bibliographystyle{ws-rv-van}

\begin{thebibliography}{99}

\bibitem{Ehrentreich} Ehrentreich, N. Agent-Based Modeling: The Santa Fe Institute Artificial Stock Market Model Revisited. Springer, 2007.

\bibitem{Gilbert} Gilbert, N. Agent-Based Models. Sage Publications, 2007.

\bibitem{Jegadeesh} Jegadeesh, N., Titman, S. Returns to Buying Winners and Selling Losers: Implications for Stock Market Efficiency. \textit{Journal of Finance} 48:1, 65--91, 1993.

\bibitem{Maymin2009} Maymin, P.Z. Regulation Simulation. \textit{European Journal of Finance and Banking Research} 2:2, 1--12, 2009.

\bibitem{Maymin2011a} Maymin, P.Z. The Minimal Model of Financial Complexity. \textit{Quantitative Finance }11:9, 1371--1378, 2011.

\bibitem{Maymin2011b} Maymin, P.Z. Markets are Efficient If and Only If P=NP. \textit{Algorithmic Finance} 1:1, 1--11, 2011.

\bibitem{Maymin2012} Maymin, P.Z.; Lim, T.W. The Iron Fist vs. the Invisible Hand: Interventionism and libertarianism in environmental economic discourses, \textit{World Review of Entrepreneurship, Management and Sustainable Development} 8:3, 358-374, 2012. 
\bibitem{Mehra} Mehra, R., Prescott, E.C. The equity premium: A puzzle. \textit{Journal of Monetary Economics }15:2, 145--161, 1985.

\bibitem{Wilensky} Wilensky, U. NetLogo. \url{http://ccl.northwestern.edu/netlogo/}. Center for Connected Learning and Computer-Based Modeling, Northwestern University. Evanston, IL, 1999.

\bibitem{Wolfram} Wolfram, S. \textit{A New Kind of Science.} Wolfram Media, 2002.

\bibitem{Wolframessay} Wolfram, S. Informal essay: Iterated finite automata. \url{http://www.stephenwolfram.com/publications/recent/iteratedfinite/}, 2003

\bibitem{Zenil} Zenil, H., Delahaye, J.-P. An Algorithmic Information Theoretic Approach to the Behaviour of Financial Markets. \textit{Journal of Economic Surveys} 25:3, 431--463, 2011.

\end{thebibliography}

\end{document}